\documentclass[structabstract]{aa}  
\usepackage{graphicx}
\newcommand{\ms}{M$_{\odot}$}
\newcommand{\zs}{Z$_{\odot}$}

\begin{document}

   \title{Relative frequencies of supernovae types: dependence on host galaxy magnitude, galactocentric radius and local metallicity }

   \author{S. Boissier
          \inst{1}
          \and
          N.Prantzos\inst{2}
%\fnmsep\thanks{Just to show the usage
%          of the elements in the author field}
        }

   \institute{Laboratoire d'Astrophysique de Marseille, OAMP, 
     Universit\'e Aix-Marseille \& CNRS UMR6110,
38 rue Fr\'ed\'eric Joliot Curie, 13388 Marseille cedex 13, France;
              \email{samuel.boissier@oamp.fr}
   \and
Institut d'Astrophysique de Paris, UMR7095 CNRS, Univ. P. \& M. Curie, 98bis Bd. Arago, 75104 Paris, France;
         \email{prantzos@iap.fr}
             }

   \date{Submitted, 2008}

\titlerunning{SN types: role of the metallicity}

%\authorrunning{}

% \abstract{}{}{}{}{} 
% 5 {} token are mandatory
 
  \abstract
  % context heading (optional)
  % {} leave it empty if necessary  
   {Stellar evolution theory suggests that the
   relationship between number ratios of supernova (SN) types and metallicity 
    holds important clues as to the nature of the progenitor stars (mass, metallicity, rotation, binarity,
    etc).}
  % aims heading (mandatory)
    {We investigate the metallicity dependence of number ratios of various SN types, using a large 
    sample of SN along with information on their radial position in, and magnitude of, their host galaxy.}
  % methods heading (mandatory)
  { We derive typical galaxian metallicities (using the well known
    metallicity-luminosity relation) and local metallicities, i.e. at
    the position of the SN; in the latter case, we use the empirical
    fact that the metallicity gradients in disk galaxies are
    $\sim$constant when expressed in dex/R$_{25}$.}
  % results heading (mandatory)
  { We confirm a dependence of the
    $N(Ibc)/N(II)$ ratio on metallicity; recent
    single star models with rotation and binary star models with no
    rotation appear to reproduce equally well that metallicity
    dependence. The size of our sample does not allow significant
    conclusions on the $N(Ic)/N(Ib)$ ratio. Finally, we find an
    unexpected metallicity dependence of the ratio of
    thermonuclear to core collapse supernovae, which we interpret in
    terms of the star formation properties of the host galaxies.}
  % conclusions heading (optional), leave it empty if necessary 
   {}


   \maketitle
%
%________________________________________________________________

\section{Introduction}

The classification of supernovae in various types (II, Ib, Ic, Ia) is
based on the presence of various features in their spectra: the
presence or absence of hydrogen distinguishes SNII from SNI, while
within the SNI family, the presence of Si lines caracterizes SNIa and
the presence of He lines distinguishes SNIb from SNIc (see e.g. Hamuy
2002, Turatto 2003 and references therein).  While SNIa are observed
in all types of galaxies (ellipticals, irregulars, spirals), SNIb and
Ic (collectively called SNIbc in this work including SN with
types Ib, Ic and Ib/c), as well as SNII are only observed in star
forming regions of spirals and irregulars; for that reason, they are
thought to originate from massive stars, as a result of the
gravitational collapse of their Fe cores (gravitational SN or core
collapse SN, CCSN in the following).  Various schemes have been
proposed to relate each one of those types to a progenitor star,
either within the framework of single star models (e.g.  Heger et al.
2003, Maeder and Meynet 2004) or invoking binary star evolution (e.g.
Eldridge et al. 2008).

A comprehensive summary of our current understanding of the various
CCSN types is provided in Fryer et al. (2007). Among the various
factors affecting the surface chemical composition of a massive star
at the time of its explosion, mass, metallicity, rotation and binarity
appear to play an important role, while the potential impact of others
(e.g. magnetic fields) is not sufficiently explored yet. The extent to
which each one of the aforementioned factors (as well as their
combined action) determines the mass lost by the star and its final
surface composition is the subject of intense theoretical and
observational investigation.  In the framework of single star models,
it is believed that, for a given metallicity, the lowest mass
progenitors of CCSN give rise to SNII, stars more massive than some
(yet poorly known) limit $M_{Ib}$ give rise to SNIb and stars more
massive than $M_{Ic}$ end as SNIc.  In principle, a lower metallicity
increases both $M_{Ib}$ and $M_{Ic}$, but its effect may be
compensated by rotation, which mixes nucleosynthesis products from the
core to the surface.

In one of the the first attempts to determine empirically the role of
metallicity in shaping the various CCSN types, Prantzos and Boissier
(2003, PB03 in the following) studied the relationship between the
number ratio of SNIbc to SNII $N(Ibc)/N(II)$ vs. the typical
metallicity of the host galaxy.  In the absence of relevant
metallicity measurements, they used the galaxian luminosity as a proxy
for metallicity, taking advantage of the well known
metallicity-luminosity relation.

Assuming reasonable values for the slope of the IMF
(in the -1.30 to -1.70 range), PB03 used the observed global
ratio of $N(Ibc)/N(II)$ to derive $M_{Ibc}\sim$20-24 \ms. 
%%%%%% REVISION 1 
This mass is comparable to the maximum mass for type II-P SN found by
Smartt et al. 2009 in their volume limited sample: 16.5 $\pm$ 1.5
M$_{\odot}$ for a Salpeter IMF (and up to 22 M$_{\odot}$ for a steeper
IMF). PB03
%%%%%%%%%
argued that this value is much lower than the one provided by
non-rotating models of single star evolution, even at high metallicity
($>$30 \ms, e.g. Heger et al. 2003). They suggested instead that
rotating models, such as the solar metallicity models of Meynet and
Maeder (2003) offer an appealing solution.
Furthermore, PB03 established a correlation of $N(Ibc)/N(II)$ with metallicity (albeit with large error bars) 
and they attributed it to the fact that $M_{Ibc}$  (the limiting mass between SNII and SNIbc)
decreases with metallicity Z. In the absence of
relevant models at lower metallicities at that time, PB03 deduced the
required $M_{Ibc}$=f(Z) relationship, in order to fit the observational
data (again, for reasonable values of the IMF slope). Subsequent
models of low Z massive rotating stars (Maeder and Meynet 2004)
confirmed relatively well those quantitative predictions.  Recently,
similar observational results have been obtained for the $N(Ibc)/N(II)$ vs Z
relation by \cite{prieto08}, who used not the mass-metallicity
relation but directly measured metallicities of a smaller sample of
host galaxies from the SDSS data base.  In the meantime, Eldridge et
al. (2008) proposed non-rotating models for binary star evolution, the
results of which reproduce also satisfactorily the $N(Ibc)/N(II)$ vs
metallicity relation. We shall discuss those theoretical developments in Sec. 5.

In this work, we investigate the relation between metallicity and number ratios 
of various SN types with a much larger SN sample than PB03 (Sec. 2).
We extend our study to the ratios of $N(Ic)/N(Ib)$ and, for the
first time, to  $N(Ia)/N(CC)$. Furthermore, we derive not only global (typical) galaxian
metallicities (Sec. 3) but also local ones, i.e. at the position of the SN within its host galaxy. For that purpose we
use the fact that metallicity gradients apear to have a constant value when expressed
in dex/R$_{25}$\footnote{The
  R$_{25}$ radius is the radius of the 25 mag arcsec$^{-2}$ B-band
  isophote.} (Sec. 4). 
We are thus able to establish statistically significant corelations with both global and local metallicity of 
$N(Ibc)/N(II)$  and, somewhat surprisingly, of  $N(Ia)/N(CC)$. We discuss those
findings in Sec. 5 and we summarize our work in Sec. 6.

\section{The supernova sample}

We use a recent version of the Asiago Supernova Catalogue (presented
in Barbon et al.\,1999) to obtain information on a large number of
supernovae and their host galaxies. This information concerns the SN
type, magnitude (usually the discovery magnitude) and relative
distance to the galactic center, the galaxy type and various
parameters, like position angle, inclination, R$_{25}$ radius
  and heliocentric radial velocity $V_{HEL}$. We use the LEDA
database (Paturel et al. 2003) to obtain for each galaxy the B-band
absolute magnitude $M_B$.
From this list, we keep only galaxies with morphological types
corresponding to spirals (S0 to Sd) and irregulars (Irr) as we are
interested in the ratio of the various supernovae types occuring 
in star forming galaxies.

\begin{figure}
   \centering
   \includegraphics[width=8.5cm]{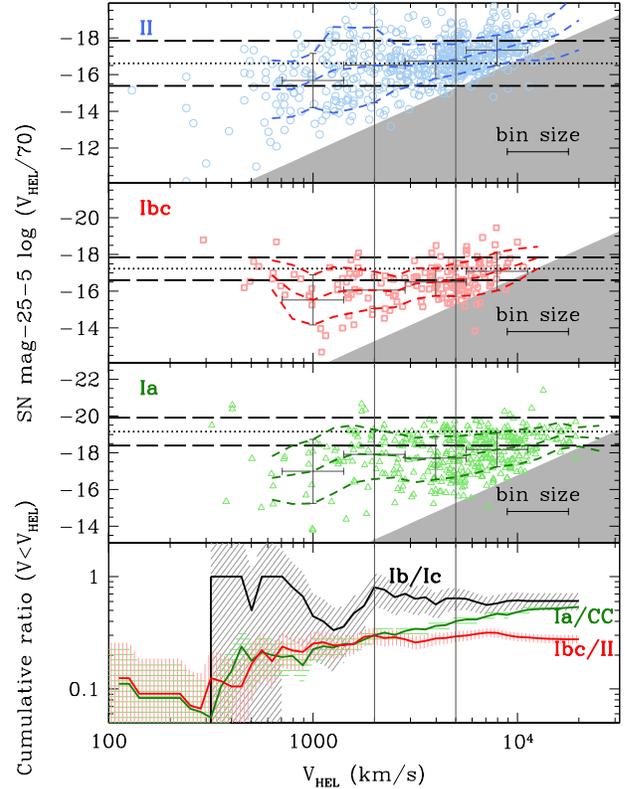}
   \caption{{\it Top} three panels: Supernovae absolute magnitude
     (adopting a basic Hubble's law with $H_{0}=70 km/s Mpc^{-1}$) as
     a function of the heliocentric velocity, for SN of type II, Ibc
     and Ia, respectively (from top to bottom).  The curves show the
     average ({\it solid}) and 1 $\sigma$ dispersion ({\it dashed})
     within bins of size indicated in the bottom-right part of the
     figure (only points separated by a distance larger than the bin
     size are independent in these curves, see text).  Independent
     points with error bars are also displayed in each panel.  The
     {\it dotted} (and {\it long-dashed}) lines show the average
     magnitude (and its dispersion) given in \cite{richardson02}.
     Almost no SN are found in the shaded area (apparent magnitudes
     fainter than 19) because it corresponds roughly to the limiting
     magnitude of the sample (see text).  {\it Bottom:} Cumulative
     ratios of SN types vs heliocentric velocity. {\it All panels:}
     the 2 vertical lines indicates the two velocity limits for which
     we present results : 5000 km/s (excluding distant galaxies for
     which we start missing large numbers of type II SN), and the more
     conservative 2000 km/s.}
         \label{FigBias1}
   \end{figure}

Because of differences in their intrinsic luminosities, various
supernovae types can be detected at different distances. In
particular, SN Ia are intrinsically brighter than core collapse SN and
can be detected further away.  Fig. \ref{FigBias1} (three top panels)
shows the SN magnitude as a function of V$_{HEL}$ in our sample. 
%% REV 1 
We are aware that these SN magnitudes are not perfectly well
determined, and not always comparable to each other (discovery vs peak
magnitude, different photometric bands, no actual distance) but,
overall, they give a rough idea of the brightness of the event.
The curves indicate the average and $\pm$ 1-$\sigma$ values. They are
obtained by computing statistics within the indicated bin size at
many points along the x-axis; the distances between those points is
less than the bin size, i.e. the points are not independent. We also
include four independent points along the curves with errorbars.
The grey shaded area indicates the absolute magnitude for SN fainter
than 19 magnitudes, where very few SN are found. This value
corresponds roughly to the limiting magnitude of SN surveys,
especially the LOSS SN Survey (van den Bergh et al. 2005) providing a
large number of SN each year. Although the Asiago catalogue does not
procure a very well defined sample, we can consider that 19 is our
approximate limiting magnitude.
Since the various SN types do not have the exact same absolute
magnitude, we have to ensure that within the volume we use, we do not
miss some types while detecting others. In order to check this, we use
the absolute magnitudes (and observation dispersion) given by
\cite{richardson02} for various types: -16.61 (1.23) for SNII; -17.23
(0.62) for SNIbc; -19.16 (0.76) for Ia. These values are given in
their Table 1 and correspond respectively to normal SN II-P, normal
Ibc and normal Ia. II-P is the most common type among SNII. Other
types (IIn, IIL) are on average brighter, thus if 
there is no bias against II-P, there should be no bias against SNII either.
%%%%%%
We note that \cite{richardson02} find evidence for a bimodal
distribution of the magnitudes of SNIbc, with a brighter component
than the normal one. Here again, if we choose our sample in such a way
to include all normal SNIbc, we should not be biased against the
brighter ones.
As long as the average $\pm$1$\sigma$ values of \cite{richardson02} do
not cross the grey-shaded area, the majority of the SN should be
present in the catalogue. Fig. 1 thus show that we may miss
significant amount of events for SNII, SNIbc, SNIa respectively beyond
$V_{HEL} \sim$ 5000, 9000, 20000 km/s.
Thus, in order to avoid biasing our sample against or in favor of one
of the SN subtypes, we have to cut it at $V_{HEL} < $ 5000 km/s.
\cite{richardson06} produced absolute magnitudes for Ib and Ic,
respectively -17.98 (scatter of 0.46), and -18.14 (0.48) for the
so-called ``normal'' ones. These values are similar and slightly
brighter than the ``normal Ibc'' taken from \cite{richardson02}, so
SNIb and SNIc should not be biased against each other, neither against
SNII adopting $V_{HEL} < $ 5000 km/s.
%%%
Note that here and in the rest of the paper, SN classified as Ib/c are
included when computing $N(Ibc)$, but are omitted when computing the
ratio $N(Ib)/N(Ic)$.
%%%%%%%%%

%%%%% REVISION 2 %%%%%%%%%%5
It can be seen in Fig. 1 that our SNIbc and SNIa are on
  average fainter than the absolute magnitude given by
  \cite{richardson02}. This is likely due to the fact that no
  extinction corrections are applied to the Asiago values which
  correspond moreover to discovery magnitudes that may differ from
  peak magnitudes (the effect is less important for SN II-P which have
  about constant magnitudes for longer time than Ia or Ibc, making
  peak and discovery magnitudes closer to each other).
%%%%%%%%%%%

Because the Asiago catalogue is quite inhomogeneous and in view of the
afore mentionned magnitudes limits, we checked the
robustness of our results by performing our analysis for samples
defined in a slightly different ways. Especially, we adopted a more
conservative approach, defining a sample with $V_{HEL} < $ 2000 km/s.
This sample should be less affected by any residual bias but suffers
from lower statistics.
  %%% 
We also checked the issue that older SN could be misclassified by
performing our analysis with the restriction of taking only events
that occurred during and after the year 2000 at the price of a smaller
number of events and larger error-bars. This affects mostly the trend
found for $N(Ia)/N(CC)$ as we will discuss in sections 3.2, 4.1, 4.2.

\cite{richardson02} note that about 20\% of SN may be sub-luminous
(with large uncertainties on this number). Of course, if the same
fractions applies to all subtypes, our results would be unchanged.
However, we should not consider results to be very robust whenever the
obtained trends could be ascribed to a variation in the number of a
sub-type by this amount.

Another possible source of bias in SN host studies is the
\cite{shaw79} effect, i.e.  the difficulty in detecting SN in inner
parts of distant galaxies, especially in photographic plates searches.
This effect depends on the SN search programs (and is thus hard to
quantify for the Asiago catalogue) however \cite{vandenbergh97}
considers that it can be neglected for $log(V_{HEL})<$
3.4. Our conservative sample ($V_{HEL} <$ 2000 km/s) should thus be
relatively unaffected.  \cite{cappellaro93} state that within the same
velocity limit ($V_{HEL} <$ 2000 km/s), 18\% of all SNe are lost in
the Asiago/Crimea searches in the overexposure of the central area of
galaxies. If all of the lost SN are from one type, the maximum error
due to this effect on a ratio is thus 18\% in the $V_{HEL} <$ 2000
km/s sample. \cite{cappellaro93} gives 23\% of lost SN for $V_{HEL} <$
4000 km/s and 35\% for their whole sample. 
%% REVISION 2 
These numbers are based on the SN missed in photographic
  searches, assuming CCD/Visual surveys did not miss any of the SN in
  the central regions. Thus, they could underestimate the real effect.
  However, we can consider they provide a first guess of the
  uncertainty on the ratios (especially those derived from the SN
  radius) due to the \cite{shaw79} effect.
A trend will be considered robust if it creates a
gradient larger than this uncertainty.

%%%%%%%%%%

In the bottom panel of Fig. 1, we show the $N(Ibc)/N(II)$,
$N(Ib)/N(Ic)$, and $N(Ia)/N(CC)$ ratios within the volume defined by
$V_{HEL}$. The intersection with the vertical lines at 5000 (2000)
km/s gives the number indicated in Table 1 for our adopted sample (and
the ``conservative'' sample).
%%%
It is interesting to note that the $N(Ibc)/N(II)$ and $N(Ib)/N(Ic)$
ratios do not depend strongly on the velocity limit adopted. This
confirms that we do not miss a large fraction of SNII vs SNIbc (and
SNIb vs SNIc) within our samples\footnote{In the worst case scenario,
  it would mean that we miss similar fractions of SNII vs SNIbc (and
  SNIb vs SNIc) when we push the velocity limit to larger values}
(this may be not so surprising as the average absolute magnitudes
differ by relatively small amount: -17.23 $\pm$0.62 vs -16.61 $\pm$
1.23 and -17.98 $\pm$ 0.46 vs -18.14 $\pm$ 0.48). The ratios
$N(Ib)/N(Ic)$, and $N(Ibc)/N(II)$ are thus quite robust, at least in
the local universe\footnote{Since the N(Ibc)/N(II) ratio depends on
  metallicity (PB03) and since cosmic metallicity decreases, on
  average, with redshift, that ratio is expected to decline with
  redshift, albeit very slowly.}.
%%%%%%%
On the contrary, the $N(Ia)/N(CC)$ ratio continuously increases when
including more distant SN (i.e. going from $V_{HEL} < $ 2000 km/s to
$V_{HEL} < $ 5000 km/s).  Although the limitation at $V_{HEL} < $ 5000
km/s should still allow us to derive meaningful results, we note that
this ratio does depend on the adopted limit and should thus be
considered as relatively uncertain.
%%%%%%%

%----------------------------------------------------------- Fig : biasgo3.ps
%
%______________________________________________________________

\begin{table}
\caption{Size of the samples and ratios}              % title of Table
\label{tab1}                                          % is used to refer this table in the text
%\centering                                            % used for centering table
\begin{tabular}{l l l }                               % centered columns (4 columns)
\hline\hline                                          % inserts double horizontal lines
SN   & Larger statistics        & Conservative  \\
Type & $V_{HEL}<$ 5000 km/s &  $V_{HEL}<$ 2000 km/s \\      % table heading
\hline                                                        % inserts single horizontal line
Ic    & 49  (41)    &   18 (15)   \\
Ib    & 32  (25)    &   15 (12)   \\
Ibc   & 98  (79)    &   43 (36)  \\
II    & 318 (239)   &  142 (96)  \\
Ia    & 166 (132)   &   56 (42)  \\
\hline                                             %inserts single line
$N(Ibc)/N(II)$  & 0.31  $\pm$ 0.04 (0.06) &    0.30  $\pm$ 0.05 (0.06 )\\
$N(Ic)/N(Ib)$   & 1.53  $\pm$ 0.35 (0.30) &    1.20 $\pm$ 0.42  (0.24) \\
$N(Ia)/N(CC)$   & 0.40  $\pm$ 0.04 (0.08) &    0.30 $\pm$ 0.05  (0.06) \\
\hline
\end{tabular}

{\it 
  Top part  : The first number is the number of SN of a 
  given type for which at least the magnitude of the host is available, 
  the second number (between parenthesis) is the number of SN for which 
  the position of the SN relative to the center of the galaxy is also known.
  Bottom part: SN ratios computed from the first of the two numbers above, with statistical error. 
  The data between parenthesis indicates what would be a 20\% error resulting from missing systematically 
  under-luminous SN from one of the sub-types; it is also the order of magnitude of the \cite{shaw79} effect for $V_{HEL} <$ 2000 km/s (see text).
}

\end{table}

Adopting V$_{HEL} < $ 5000 km/s, the catalogue provides
1038 SN, 701 occuring in star forming galaxies (spirals
and irregulars) and for which we have at least the host galaxy
magnitude. The sample we will use to study the core-collapse ratios is
then a factor 2.5 larger than the 280 CCSN used in PB03. The distribution
of SN types in this sample is given in Table 1.

The SN ratios that we obtain for the different SN types of our sample
are consistent with published values: the ratio of SNIbc to SNII
supernovae in our sample is $N(Ibc)/N(II)$=0.31$\pm$0.04, similar to
0.27 in PB03; 0.23$\pm$0.05 in Bressan et al. (2002); 0.33 in Hamuy
(2002); 0.41 in \cite{smartt08}, 0.3, 0.16 and 0.31 for respectively
S0a/b, Sbc/d and Irr galaxies in Mannucci et al. (2005).  The ratio of
thermonuclear to core collapse SN in that sample is
$N(Ia)/N(CC)$=0.4$\pm$0.04, and the same ratio is obtained from the
local universe supernovae sample (Smartt et al. 2009).  Using the
rates from \cite{mannucci05}, we obtain for that ratio the values
0.41, 0.19, and 0.34 in S0a/b, Sbc/d, and Irr galaxies, respectively;
the corresponding uncertainties, however, are very large, due to small
statistics. Our result for $N(Ic)/N(Ib)$=1.65 $\pm$ 0.32 is consistent
with the one obtained from the local universe sample of Smartt et al.
(2009): $N(Ic)/N(Ib)$=2 considering the small statistics (27 SN) in
their sample for this ratio (compensated however by a careful
  checking of the data for every SN used in their work).

   \begin{figure*}
   \centering
   \includegraphics[angle=-90,width=16cm]{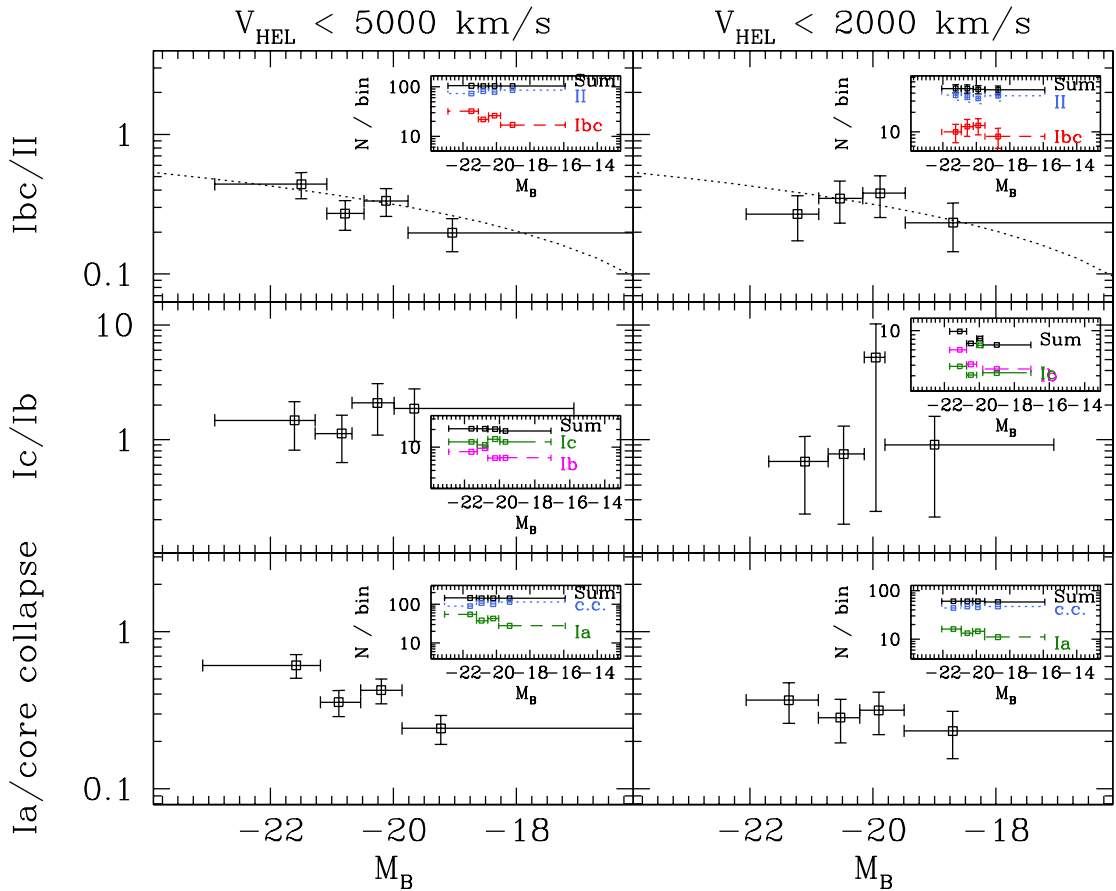}
   \caption{From {\it top } to {\it bottom}: number ratios of
     $N(Ibc)/N(II)$, $N(Ic)/N(Ib)$ and $N(Ia)/N(CC)$, as function of
     galaxian blue magnitude $M_B$ for the conservative
       $V_{HEL}<$ 2000 km/s sample ({\it right}) and the $V_{HEL}<$ 5000
       km/s sample ({\it left}, better statistics). The inserts in each
     panel display the corresponding numbers of each SN type and are
     chosen such as each $M_B$ bin has the same total number of SN of
     the types involved.  The {\it dotted} curve in the upper panel is
     the fit to the data of PB03, which had much larger error bars.  }
\label{Fig:SNratios_vs_magn}
   \end{figure*}

\section{Dependence of SN type ratios on global  galaxy properties}
\label{secMAG}

Using a sample of 280 CCSN from an earlier version of the Asiago
catalogue, PB03 found that the $N(Ibc)/N(II)$ ratio has an average
value of $\sim$0.30, while it increases with host galaxy magnitude.
They interpreted the latter as an effect of the global galaxian
metallicity (increasing with galaxian luminosity) on the masses of the
precursors of the CCSN sub-types: as metallicity increases, the
stellar enveloppe is more easily lost and lower mass stars may become
SNIbc, increasing thus the $N(Ibc)/N(II)$ ratio. 
Assuming all CCSN are produced from single stars, and that the
physical reason for a star to explode as a SNIc, SNIb, SNII is only
its initial mass (see the discussion in section 5 for other
possibilities), the $N(Ibc)/N(II)$ ratio
is expressed as
\begin{equation}
{{N(Ibc)}\over{N(II)}}  \ = \ {{\int_{M_{Ibc}}^{M_{Up}} \Psi(t-\tau_M)\Phi(M)dM} \over
{\int_{M_{II}}^{M_{Ibc}}\Psi(t-\tau_M )\Phi(M)dM}}
\label{eqNIbc_overNII}
\end{equation}
where $\Psi(t)$ is the star formation rate at time $t$, $\tau_M$ the
lifetime of star of mass $M$, $\Phi(M)$ the stellar initial mass
function (IMF), and $M_{Up}$, $M_{II}$ and $M_{Ibc}$, respectively,
the upper mass limit of the IMF (around 100 \ms), the lower mass limit
for a star to explode as SNII (around 8 \ms, see e.g. Smartt et al.
2009, Anderson \& James 2008) and the lower mass limit for a star to
explode as SNIbc.
Under the crucial assumption of
stationarity, whereby the progenitor lifetimes of both classes of SN
(i.e. SNII and SNIbc) are short compared to the duration of the
corresponding star formation episodes, $\Psi$ can be taken out of the
integral in Eq. (1) and cancelled (since the progenitors of both
classes of SN have similar lifetimes, during which $\Psi$ varies very
little). Then, Eq. (1) reads:
\begin{equation}
{{N(Ibc)}\over{N(II)}} \ = \ {{\int_{M_{Ibc}}^{M_{Up}} \Phi(M)dM}\over{\int_{M_{II}}^{M_{Ibc}} \Phi(M)dM}}
\label{eqNIbc_overNII2}
\end{equation}
i.e. the $N(Ibc)/N(II)$ ratio is a function of the slope of the IMF
and of $M_{Ibc}$. For a given IMF, if $M_{Ibc}$ increases with
metallicity, the ratio given by Eq. (2) will obviously decrease and
the same is true if one replaces $N(Ibc)/N(II)$ with $N(Ic)/N(Ib)$ and
$M_{Ibc}$ with $M_{Ic}$.  Note that \cite{anderson08} recently
suggested such a hierarchy of limiting masses between SNII, SNIb, SNIc
on the basis of the association of SN with regions of recent star
formation traced by H$\alpha$ emission. Their findings indicate
$M_{II} \sim$ 7.8 \ms. \cite{kelly08} found that SNIc occur in the
brightest regions of their host, where the most massive stars probably
form, also suggesting that SNIc results from the explosion of the most
massive stars.

\subsection{The  $N(Ibc)/N(II)$ and $N(Ic)/N(Ib)$ ratios}

In this section we first repeat the analysis as PB03 for the ratios of
CCSN subtypes with our larger sample.  In Fig. 2, the top panels show
the $N(Ibc)/N(II)$ ratio vs $M_B$. Four bins in $M_B$ with $\sim$100
CCSN in each one are constructed and the corresponding $M_B$ value is
taken as the median value of $M_B$ in each bin. 
In can be seen in the insert panel that in bins of constant CCSN
numbers, $N(Ibc)$ increases for brighter galaxies ($N(II)$ decreasing
by the same amount),
%Clearly, the number of SNII decreases for brighter galaxies and the
%number of SNIbc increases (see insert panel) 
making the $N(Ibc)/N(II)$ ratio to increase with galaxian luminosity.
The resulting $N(Ibc)/N(II)$ vs $M_B$ relation is quite similar to the
one obtained in PB03 (dotted curve), with smaller vertical error
bars, reflecting the larger size of the new sample.
We thus confirm the original result of PB03 between the $N(Ibc)/N(II)$
ratio and {\it global} galaxian metallicity, which is also supported
by the study of Prieto et al.  (2008). 
%%%%%%%%
The only limitation to this conclusion is obtained with the
  conservative sample for which the relation is rather flat. The main
  difference come from the bin corresponding to the brightest galaxies
  showing low values with respect to the $V_{HEL}<$5000km/s sample and
  the PB03 fit (errorbars are however rather large).
%%%%%%%%
We shall return to the
interpretation of that result in Sec.  5, after presenting in Sec. 4
the results of our study concerning that same ratio as a function of
{\it local} metallicity.

The $N(Ic)/N(Ib)$ ratio (middle panel of Fig. 2) has considerably
larger uncertainties than the $N(Ibc)/N(II)$ ratio, because of poorer
statistics.  Indeed, each one of the four bins contains $\sim$ 8 SNIb
and $\sim$12 Ic (insert in middle panel), for an average ratio of
$N(Ic)/N(Ib)\sim$1.6  That ratio shows no clear variation with $M_B$.
At first sight, this appears to indicate a situation opposite to the
case of the $N(Ibc)/N(II)$ ratio. We shall see, however, in Sec. 4,
that the situation is different when $N(Ic)/N(Ib)$ is expressed as a
function of local metallicity. This ``puzzling'' behaviour is further
discussed in Sec. 4 and attributed to small number statistics.

\subsection{The  $N(Ia)/N(CC)$ ratio }

In this section, we extend our study to the $N(Ia)/N(CC)$ ratio as a
function of host galaxy M$_B$.  As can be seen in the bottom panels of
Fig. 2, there is also a clear trend in that case, with brighter
galaxies hosting proportionally more SNIa than CCSN. To our knowledge,
it is the first time that such a correlation is put in evidence. In
view of the numbers of SN involved (see insert panel) we feel that
this result is rather robust, contrary to the case of $N(Ic)/N(Ib)$.
We note that we still find such a correlation for the conservative
limit $V_{HEL} < $ 2000 km/s (right panel) although it is
less steep in that case.  On the other hand, taking only recent SN
from the sample (trying to avoid misclassifications), the correlation
gets steeper. The dependence of $N(Ia)/N(CC)$ on the magnitude is thus
relatively uncertain in absolute value, however the fact that we do
find a correlation with magnitude in every one of our tests let us
think that it is real.

That result can be translated in terms of metallicity, with larger
$N(Ia)/N(CC)$ ratios found in more metal-rich galaxies. We shall see
indeed in Sec. 4 that this result also holds when the local
metallicity is considered. However, we think that, contrary to the
previous case, there is no causal relationship here, i.e. metallicity
does not affect in any way the $N(Ia)/N(CC)$ ratio. Instead, it is the
mass of the host galaxy which affects that ratio, in a way that can
most easily be seen as follows.

The CCSN rate of a galaxy is proportional to the star formation rate $\Psi$:
\begin{equation}
N(CC) \ \propto \ \Psi 
\end{equation}

The SNIa rate is more difficult to evaluate, since thermonuclear
supernovae may result from stars of all ages, not just the young ones.
Scannapieco and Bildsten (2005)  introduced a useful parametrization, describing the SNIa
rate as a sum of two terms, one depending on the current star
formation rate and the other on the total stellar mass $M_*$ (i.e. the
time integrated star formation rate). That parametrization (adopted by
e.g. Sullivan et al. 2006, Aubourg et al.  2008) fits satisfactorily
most available data and we adopt it here:
\begin{equation}
N(Ia) = \alpha \ \Psi  \ + \ \beta \ M_*
\end{equation}
where $\alpha$ and $\beta$ are constants. Thus, the required ratio can be written as
\begin{equation}
{{N(Ia)}\over{N(CC)}} \ = \ A  + \ B \ {{M_*}\over{\Psi}}
\end{equation}
Now, it is well known that larger galaxies have, on average, smaller
specific star formation rates $\Psi$/M$_*$ because of their smaller
gas fractions (e.g. Boissier et al. 2001; Boissier and Prantzos 2000
and references therein), i.e. the ratio $M_*/\Psi$ is, on average, an
increasing function of galaxy mass and luminosity. Thus, both the
metallicity of a galaxy and its $N(Ia)/N(CC)$ ratio increase with its
mass. This explains, at least qualitatively, the correlation found in
the bottom panels of Fig. 2.

\section{Dependence of SN type ratios on local  galaxy properties}

   \begin{figure*}
   \centering
   \includegraphics[angle=-90,width=16cm]{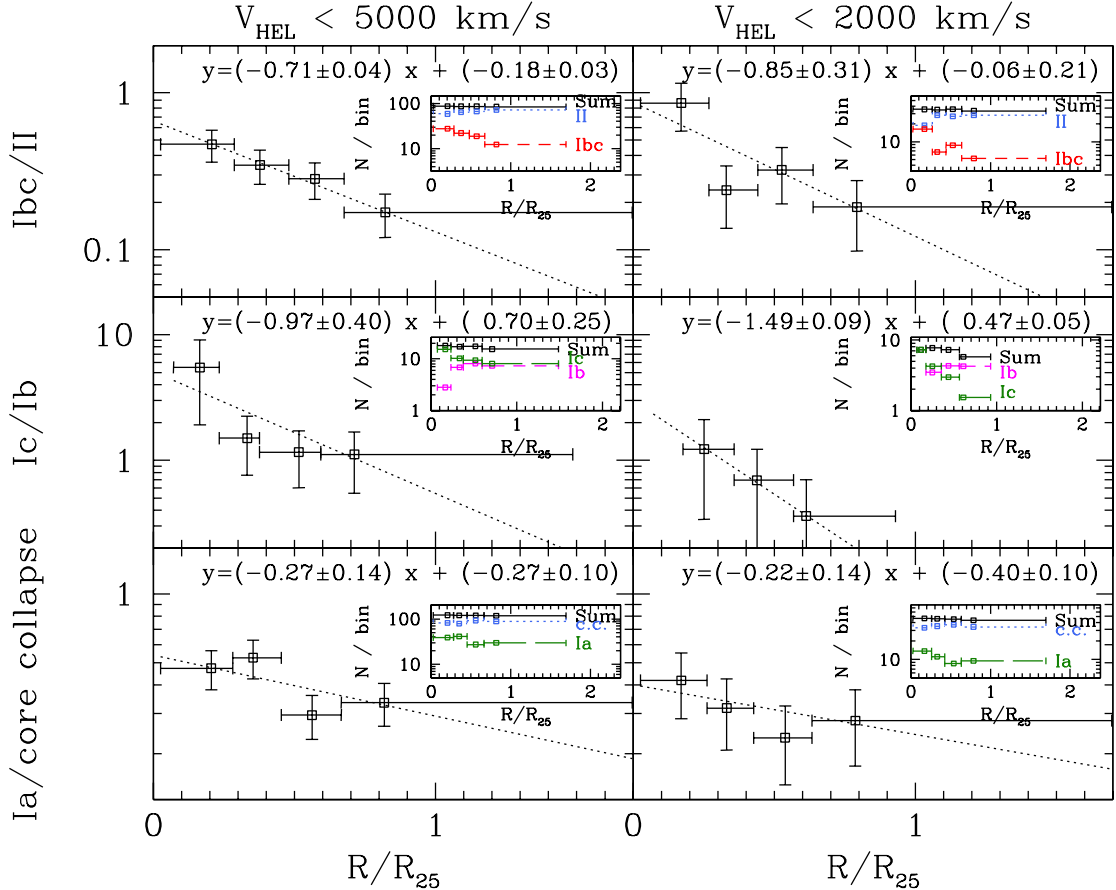}
   \caption{Same thing as in Fig. 2, but this time as a function of
     galactocentric radius, expressed in units of R$_{25}$. }
         \label{Fig:SNratios_vs_rad}
   \end{figure*}

PB03  used the absolute
magnitude of the host galaxy as a proxy for its global metallicity, based on the well established magnitude-metallicity relationship.
However, disk galaxies are known to exhibit metallicity gradients 
(e.g. Henry and Worthey 1999, Zaritsky et al. 1994, van Zee et al. 1998). 
If metallicity affects indeed the ratios of SN types, a radial effect
should also be found.
\cite{hakobyan08} has indeed shown that the radial distributions of SNIbc and SNII
are different, with more SNIbc found at smaller radii than SNII (using a sample extracted
from the Asiago catalogue also, but with a different selection than ours),
A similar analysis was made by  \cite{vandenbergh97}, for a smaller sample of 156 SN.

\subsection{Dependence on galactocentric distance}

In this section we investigate whether such a radial effect is seen in
our sample, using the supernovae for which we can compute a
galactocentric radius.  This is possible when the catalogue provides
the offset of the SN, the position angle and inclination of the host
galaxy, as well as its $R_{25}$ radius, which is needed to normalize
the results. From these parameters, we compute the distance between
the SN position and the center of the galaxy, within its plane, that
we will call the galactocentric radius of the SN.  Note that we
de-project the minor axis simply as $b/a$=cos$(inclination \ angle$),
and we do not use the galaxies which are almost edge-on (inclination
larger than 80 degrees).  These restrictions reduce the size of the
usable sample for this part of the study, but only moderately
(see Table 1) allowing us to work with decent statistics.

Fig. 3 displays the same ratios as Fig. 2, this time as a function of
galactocentric radius.  A clear trend is observed in the case of
$N(Ibc)/N(II)$. SNIbc are found at smaller normalised radius than
SNII, in agreement with \cite{hakobyan08}.  \cite{vandenbergh97}
already suggested from his small sample that SN Ibc were more
concentrated toward the central part of their host galaxy than SNII.
As mentioned in section 2, the \cite{shaw79} effect should be lower
than 35\%, but the difference between the inner and outer bins is much
larger (about a factor 2.5). Actually adopting the conservative
sample (right panel), the \cite{shaw79} effect should be even smaller, and we still
find a trend (actually, even stronger: slope of -0.85 instead of
-0.71). Thus we believe this trend to be unaffected by this source of
bias.
%%%%%%%
The easiest way to
interpret this observation is in term of metallicities: larger
metallicities are found in inner parts of galaxies, leading to a lower
limiting mass for type Ibc supernovae.  We shall quantify the effect
in Sec. 5.2 in terms of local metallicities, showing that it is
consistent with what we obtained in Sec. 4 by using global
metallicities.

The clear trend obtained in the case of $N(Ic)/N(Ib)$ (middle panels
in Fig. 3) is rather surprising, in view of the results of Sec. 2:
$N(Ic)/N(Ib)$ apparently increases with decreasing galactocentric
radius (i.e. with increasing metallicity) while no variation with
$M_B$ is seen in Fig 2.
We note that with a difference between inner and outer bins of a
factor about 5, here again, we cannot ascribe the observed trend to
the \cite{shaw79} effect (if we adopt the conservative sample, we do
not have any SN Ib in the inner most bin, but the trend in the three
other bins is stronger).
%%%
We attribute the striking difference between the trend with magnitude
and radius to the small number statistics involved in the evaluation
of that ratio. As stressed in the beginning of Sec. 4, the assumption
of stationarity is crucial in the evaluation of the various SN ratios.
That assumption is naturally fullfilled if large numbers of SN are
involved. In that case, the formation times of the SN progenitors of
all types span the whole range of the progenitor lifetimes ($\tau_M$);
an average $\Psi$ can be used then, allowing one to pass from Eq. (1)
to Eq. (2).  However, in the case of small numbers of SN the situation
is different: if a few starbursts occured recently (less than a few
Myr ago), only the most massive of their stars had time to explode up
to now, favouring SNIc (presumably resulting from more massive stars)
over SNIb (and, for the same reasons, SNIbc over SNII). In that case,
the term $\Psi(t-\tau_M)$ in Eq. (1) does not cancel out with the
corresponding term in the denominator and may mask the effect of any
metallicity dependence of $M_{Ibc}$ (the dividing mass between SN
exploding as Ibc or II) or of $M_{Ic}$ (the dividing mass between SN
exploding as Ib or Ic). We shall see in the next section that the
radial trend of $N(Ic)/N(Ib)$ found here translates directly into a
local metallicity trend, but because of low number statistics it is
impossible to draw meaningfull conclusions.

Finally, the bottom panels of Fig. 3 displays the ratio of
thermonuclear to core collapse supernovae $N(Ia)/N(CC)$.  The ratio
appears to increase in the inner galaxian zones. 
%Since the statistics
%(see insert in that panel) is similar to the case of $N(Ibc)/N(II)$
%(about 15-30 SNIa for more than 100 CCSN in each bin).  
%%%
The trend is relatively weak: with the slope and uncertainty in the
figure, a Student's t-test indicates a $~$ 20\% probability for the
null hypothesis that there is no dependence of $N(Ia)/N(CC)$ on the
radius. Very similar results are obtained for the conservative sample
($V_C <$ 2000 km/s).
Another worrying issue is that inner and outer bins are different only
by a factor $\sim$ 1.4. This is still larger than the typical
uncertainties due to the \cite{shaw79} effect or under-luminous SN,
but in combination with the large statistical error-bars, it makes
this trend less robust vs uncertainties than the other ones presented
in the figure.
%%%
Interestingly, if we keep only SN that exploded during or after the
year 2000, avoiding possible misclassifications for older SN, we find
a steeper trend with radius (slope -0.46 $\pm$ 0.06, reducing the
probability for the null hypothesis to less than 1\%). In summary,
although the trend in Fig. 3 is not very strong, it is reasonable to
believe it is real in the sense that we still find it when reducing
the size sample with various criteria aiming to improve its quality.
%%%
It is rewarding that such a trend is also expected
on the basis of the analysis made in Sec. 3.2. Indeed, the gaseous
profiles of disk galaxies vary little with galactocentric radius,
while the stellar ones much more (for instance, in the case of the
Milky Way disk the scalelength of the stellar profile is $\sim$2.5
kpc, while the one of the gas $\sim$8 kpc; see e.g. Boissier and
Prantzos 1999 for observed stellar and gaseous profiles of the MW
disk). This implies that the term $M_*/\Psi$ in Eq. (5) is expected to
increase in the inner galaxian zones and so will do the corresponding
$N(Ia)/N(CC)$ ratio. We develop further this argument analytically in
Sec. 5.3 and we illustrate it with a numerical example from the case
of the Milky Way disk.  We shall see in the next section that this
radial trend is also expressed in terms of local metallicity; however,
as already argued in Sec. (2), metallicity is not the root cause of
that effect.

\subsection{Dependence on local metallicity}

In order to probe the metallicity dependence of SN types, one may use direct or indirect
methods to determine metallicity. Direct measurements of abundances of
host galaxies of supernovae are possible only for moderately large  samples of SN
(e.g. 254 galaxies from the SDSS in Prieto et al., 2008) and these are integrated
abundances over the whole galaxy, not at the supernova position.
Measuring local metallicities, i.e. in HII regions at the immediate vicinity of
the supernova, would be ideal but it would require an enormous observational effort
before  numbers for decent statistics are obtained.
%%%REVISION1
Determinations of local metallicities from spectroscopy of neighboring
HII regions do exist (Smartt et al. 2009, Modjaz et al.  2008) but
only for relatively small number of events, not allowing the
statistical study of the various trends explored in this work.

In PB03 we used an indirect way, i.e. the well-known mass-metallicity relationship 
 to evaluate the metallicity of the host  galaxies (of late type) of CCSN. 
In this work, we use known relationships among disk galaxies in order to derive again
in an indirect (and approximate) way the local metallicity of galaxies at
the galactocentric  radius of the supernovae of our sample. 

It is known that the abundance
gradient in nearby disk galaxies has a universal value when expressed
in dex/$R_{25}$ (e.g. Henry and Worthey, 1999) of dlog(O/H)/dR$_{25}\sim$-0.6 dex/R$_{25}$. 
Prantzos and Boissier (2000) showed, with
detailed semi-analytical models of disk galaxy evolution, that this universality can indeed be 
reproduced, thus confirming an earlier suggestion of Garnett et al. (1997)
on ``homologuous'' disk evolution. Combining this empirical fact 
with the  observed luminosity-metallicity relation,
it is possible to deduce the metallicity at the vicinity of the SN from 
the luminosity of the host galaxy and the galactocentric radius of the SN (which is already evaluated in Sec. 4.1).

   \begin{figure}
   \centering
 \includegraphics[width=0.49\textwidth]{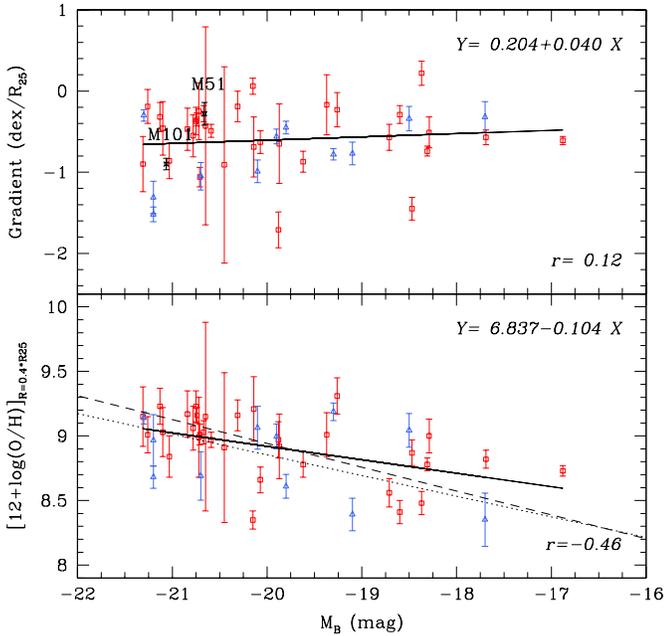}
 \caption{{\it Top:} Metallicity (oxygen) gradient per $R_{25}$ as a
   function of absolute magnitude $M_B$.  {\it Bottom:} Characteristic
   metallicity (at 0.4 $R_{25}$) as a function of $M_B$. Data in both
   panels are from \cite{zaritsky94} (squares) and \cite{vanzee98}
   (triangles).  $r$ is the correlation coefficient, and the {\it
     solid} line indicates in each case the best linear least squares
   fit (the parameters of the line are given in each panel).  In the
   top panel, the gradients recently derived by \cite{bresolin04} and
   \cite{bresolin07} for M51 and M101 are overploted. In the bottom
   panel, {\it dashed} and {\it dotted} lines indicate, respectively
   the empirical relations obtained by Tremonti et al.  (2004) and
   Garnett (2002) using respectively global metallicities (measured on
   the integrated galaxy) and metallicities at the effective
   (half-light) radius.  The latter was used as a proxy for the global
   metallicity in PB03. We apply the same approach in this paper for
   the global metallicity.}
         \label{Fig:Met_Grad}
   \end{figure}

%
%----------------------------------------------------------- Fig : biasgo4.ps
   \begin{figure}
   \centering
   \includegraphics[width=0.49\textwidth]{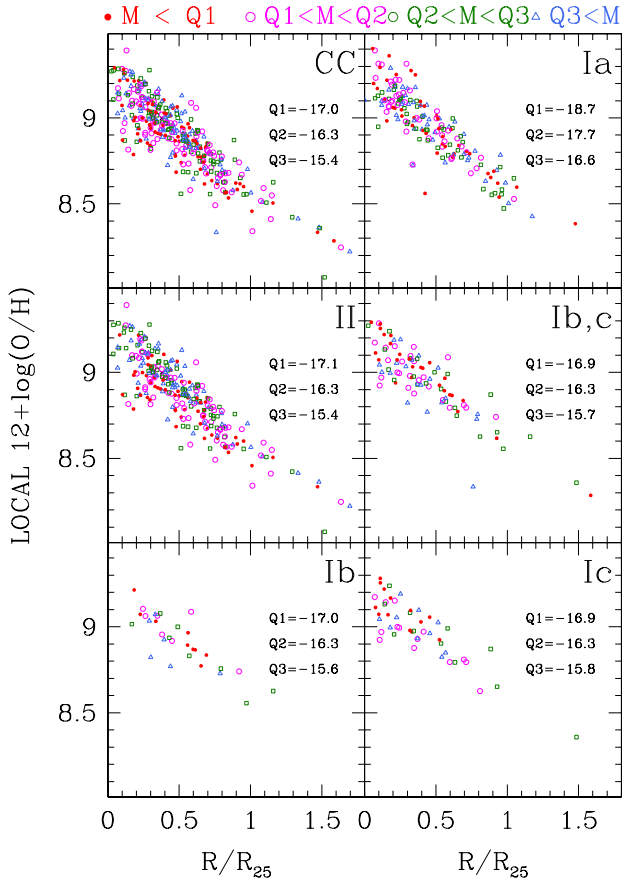}
   \caption{Metallicities of the various SN types of our sample, as a
     function of their position (expressed in $R/R_{25}$) inside their
     host galaxy. Metallicities are derived from the SN position in
     and the magnitude of the host galaxy, according to the procedure
     described in Sec. 5.2 and Eq. 6. The color/type coding is done
     according to the absolute magnitude of the SN with respect to the
     three quartiles Q1, Q2, Q3 indicated in each panel.}
\label{Fig:SNmetallicities}
   \end{figure}
%
%______________________________________________________________

For our purpose we use the data of two studies of abundance gradients
with relatively large samples of nearby galaxies : \cite{zaritsky94}
and \cite{vanzee98}.  We show in Fig.  4 (top) the abundance gradients
in dex/$R_{25}$, which display little variation (if any at all) with
absolute magnitude.  In the bottom panel of Fig. 4 is displayed the abundance
measured at 0.4 $R_{25}$ as a function of the absolute B band
magnitude $M_B$ (the metallicity-luminosity relation).  Each panel
features a linear least square fit for these relationships (solid
lines) that we adopt in this work to compute the local metallicity.
The dashed and dotted lines in the bottom panel of Fig. 4 indicate the
metallicity-luminosity relations of Tremonti et al.  (2004) and
Garnett (2002). The first one was obtained from intergrated galaxy spectra of the Sloan
Digital Sky Survey. The second one displays
the metallicity at the effective (half-light) radius and was
used in PB03 to estimate the metallicity in SN hosts without knowing
the galactocentric distance of the SN. In this paper, for consistency,
we still use it to estimate the global metallicity.  Despite using
different definitions of the ``characterisitic'' abundance, the aforementioned
relationships are very close to each other, and our results would be
marginally affected by the use of either of them.

 %, which are slightly steeper than ours,
 %  but our results are only marginally affected by that difference.

%-----------------------------------------------------------
\begin{figure*}
  \centering
  \includegraphics[angle=-90,width=0.8\textwidth]{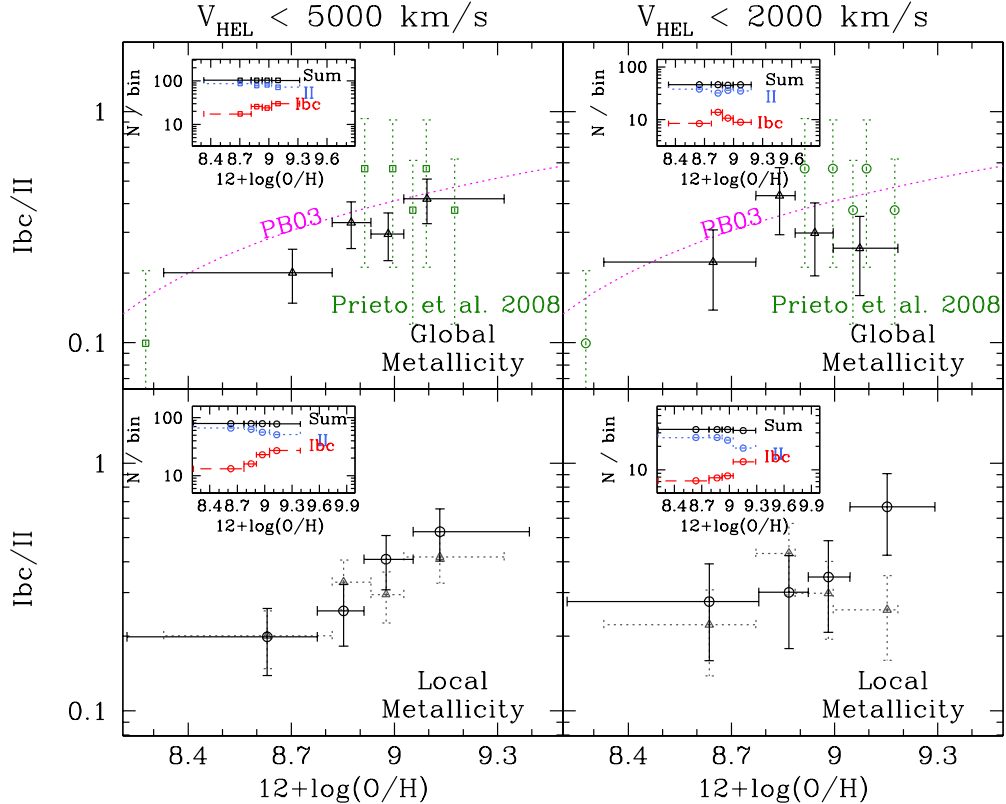}
  \caption{Symbols with {\it solid errror bars}: Number ratios of
    SNIbc/SNII as a function of global galaxian metallicity ({\it
      top}, obtained from the magnitude-metallicity relation of Fig.
    4) and of local metallicity ({\it bottom}, obtained from Eq. 6),
    for the conservative $V_{HEL}<$ 2000 km/s sample ({\it right}) and
    the $V_{HEL}<$ 5000 km/s sample ({\it left}, better statistics).
    Inserts in all panels display corresponding numbers of SN in each
    bin (chosen such as the total numbers are approximately equal).
    {\it Triangles} with {\it dotted error bars} in lower panels are
    the same as those with {\it solid error bars} in the upper panels
    and are displayed for comparison with local metallicity results.
    In the {\it upper} panels, the {\it dotted curve} is the fit to
    the data of PB03 and the symbols with {\it dotted vertical error
      bars} the data collected in \cite{prieto08}.  }
  \label{FigMet}
\end{figure*}

%-----------------------------------------------------------
\begin{figure*}
  \centering
  \includegraphics[angle=-90,width=0.8\textwidth]{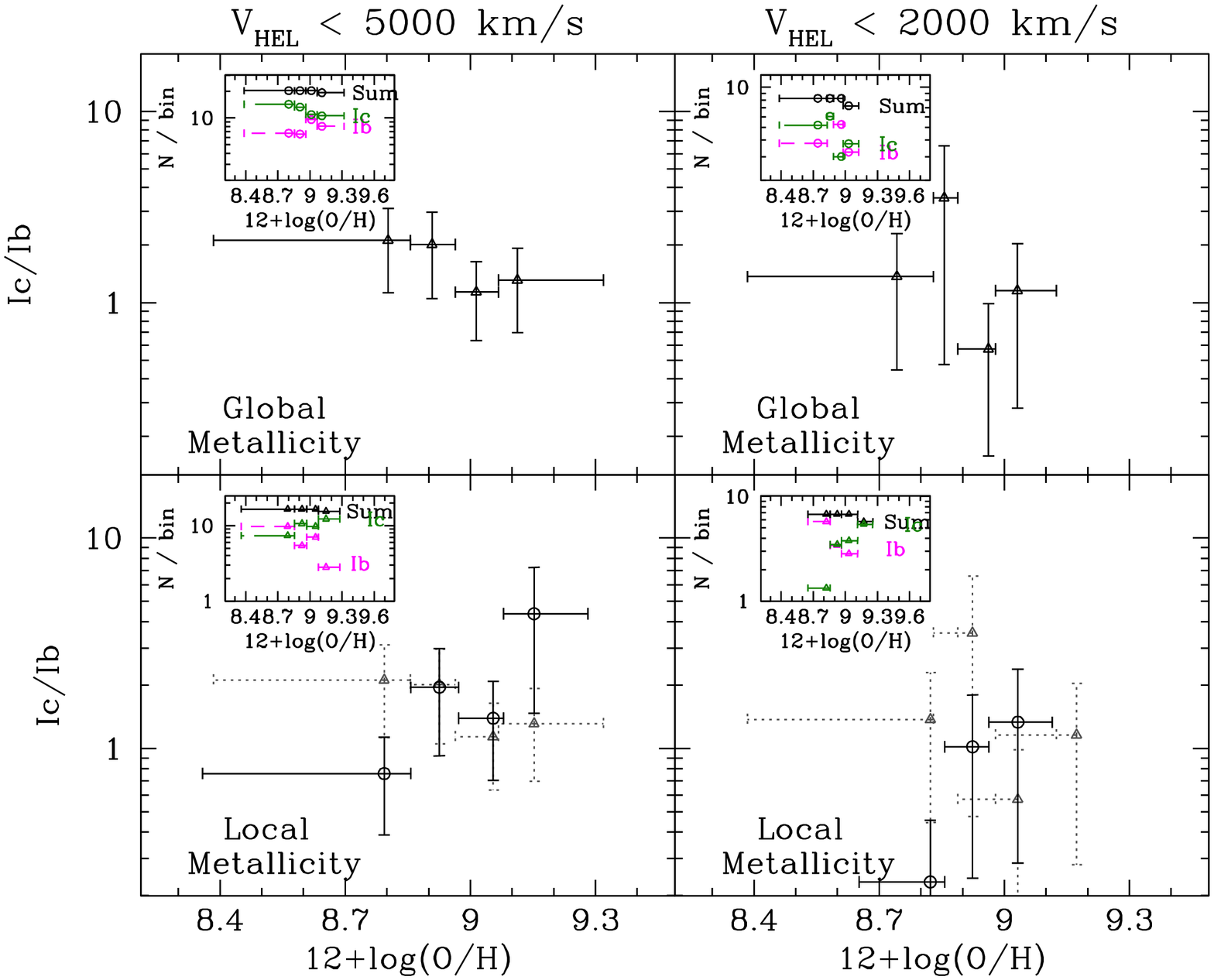}
  \caption{Symbols with {\it solid errror bars}: Number ratios of
    SNIc/SNIb as a function
    of global galaxian metallicity ({\it top}, obtained from the
    magnitude-metallicity relation of Fig. 4) and of local metallicity
    ({\it bottom}, obtained from Eq. 6), for the conservative
       $V_{HEL}<$ 2000 km/s sample ({\it right}) and the $V_{HEL}<$ 5000
       km/s sample ({\it left}, better statistics).  Inserts in all panels
    display corresponding numbers of SN in each bin (chosen such as
    the total numbers are approximately equal). {\it Triangles} with
    {\it dotted error bars} in lower panels are the same as those with
    {\it solid error bars} in the upper panels and are displayed for
    comparison with local metallicity results. }
  \label{FigMetcsurb}
\end{figure*}

Assuming that the two empirical relations (i.e. the solid lines in the
top and bottom panels) are valid for all the galaxies of our sample,
we can compute the metallicity profile (in terms of $R/R_{25}$) of
each galaxy from its luminosity. We can then evaluate easily the local
metallicity at the galactocentric radius of the supernova. Using the
numerical values of Fig. 4 (i.e the fits appearing in each panel) we
obtain:
\begin{eqnarray}
\left[ 12 + {\rm log(O/H)} \right] (R)  & = & 6.837-0.104 \ M_B  \\
                                   & + & ( \frac{R}{R_{25}} - 0.4) (0.204+0.04 \ M_B) \nonumber 
\end{eqnarray}
Notice that, although the abundance gradient has an almost universal
value in dex/$R_{25}$, our fit provides a very small trend with $M_B$,
which appears as the last term (dependence on $M_B$) in the derived
expression. We keep this term for consistency, but it is clear that,
in view of its small magnitude, it has no influence on the results.

The resulting local metallicities $O/H(R)$ as a function of normalized
galactocentric radius $R/R_{25}$ appear in Fig. 5 for all the SN of
our sample: the upper panels displays CCSN and SNIa, the middle panels
SNII and SNIbc and the bottom panels SNIb and SNIc. In all cases, the
metallicity gradient is the same, but the absolute value of the
metallicity at each normalized radius depends on the corresponding
host galaxy magnitude. It is the first time that this technique is
used in order to derive local metallicities for SN progenitors. Its
results depend obviously on how accurately the adopted {\it average}
relationships (metallicity-luminosity and gradient-luminosity) apply
to each galaxy of our sample.

With the metallicities at the position of each supernovae derived in
this way, we compute then the corresponding ratios as a function of
the local metallicity.  In Fig. 6 and Fig. 7,  we present the results for respectively the
ratios $N(Ibc)/N(II)$ and $N(Ic)/N(Ib)$ 
as a function of global metallicity (top, after the results of Sec.
4.1 and the magnitude-metallicity relationship) and of local
metallicity (bottom).

Before discussing these results, we want to mention that we performed
the same figure adopting different abundance gradients.  During the
last years, several studies have mentioned the possible errors in the
abundances obtained from strong lines as in \cite{zaritsky94} or
\cite{vanzee98}. \cite{bresolin04} and \cite{bresolin07} performed a
more detailed analysis of the gradients in M51 and M101. They found
flatter gradients than in previous studies for the same galaxies,
however their gradients are within the dispersion of those in Fig. 4.
We performed the same analysis as described above but adopting the two
values for M51 and M101 rather than our fit. We find that it does not
affect our results qualitatively. The only difference is that flatter
gradients make the trend with metallicity steeper and vice versa.

In the case of $N(Ibc)/N(II)$, the results of the $V_{HEL} <$
  5000 km/s as a function of global metallicity are consistent with
those obtained in PB03 (dotted curve) and those of Prieto et al.
(2008, vertical dotted error bars). We note the relatively good
statistics, due to the size of our sample (see insert).  The
corresponding ratios as a function of local metallicity (bottom panel)
are quite consistent with those obtained for global metallicity and
the statistics is almost equally significant. 
Looking at the $V_{HEL} <$ 2000 km/s sample, we notice that
  the trend with global metallicity is consistent with an absence of
  relationship. However, this is mostly due to the higher metallicity
  bin (the three other ones are indeed within 1 sigma of the PB03
  fit). We think that small statistics are partly responsible for this
  difference. The relation with local metallicity however (that should
  be a better estimator of the actual progenitor metallicity) is
  totally consistent with the one obtained with the $V_{HEL} <$ 5000
  km/s sample.
We conclude then that
the trend of $N(Ibc)/N(II)$ with metallicity, first identified in
PB03, is firmly established
as long as the two types Ibc and II are unbiased as a function of
$R/R_{25}$. The discussion in section 4.1 suggests that it is the
case. Also, in Fig. \ref{Fig:SNmetallicities}, the symbol/color coding
according to the SN magnitudes shows that SN in the inner galactic
regions are not apparently biased towards the brighter ones.
The case of $N(Ic)/N(Ib)$ is much less clear. Results show no trend
with global metallicity (as expected from Sec. 3.1) but they do show
such a trend as a function of local metallicity (as expected from Sec.
4.1), even if the slope and intersect are poorly constrained
  (see the error-bars and the difference between the two samples shown
  in Fig. 7)
Due to the smaller samples (see inserts) statistics is poorer in
that case, as reflected in the large error bars. The arguments
developped in Sec. 4.1 suggest that such a conflicting situation can
occur indeed: Eq. 2 does not apply and Eq. 1 (which always applies)
can produce ambiguous results, depending on the ages of the few
starbursts involved. It is then impossible to draw any conclusions
about the dependence of that ratio on metallicity; a substantially
larger sample is required for that.

% ----------------------------------------------------------- Fig : biasgo4.ps
\begin{figure*}
  \centering
  \includegraphics[angle=-90,width=16cm]{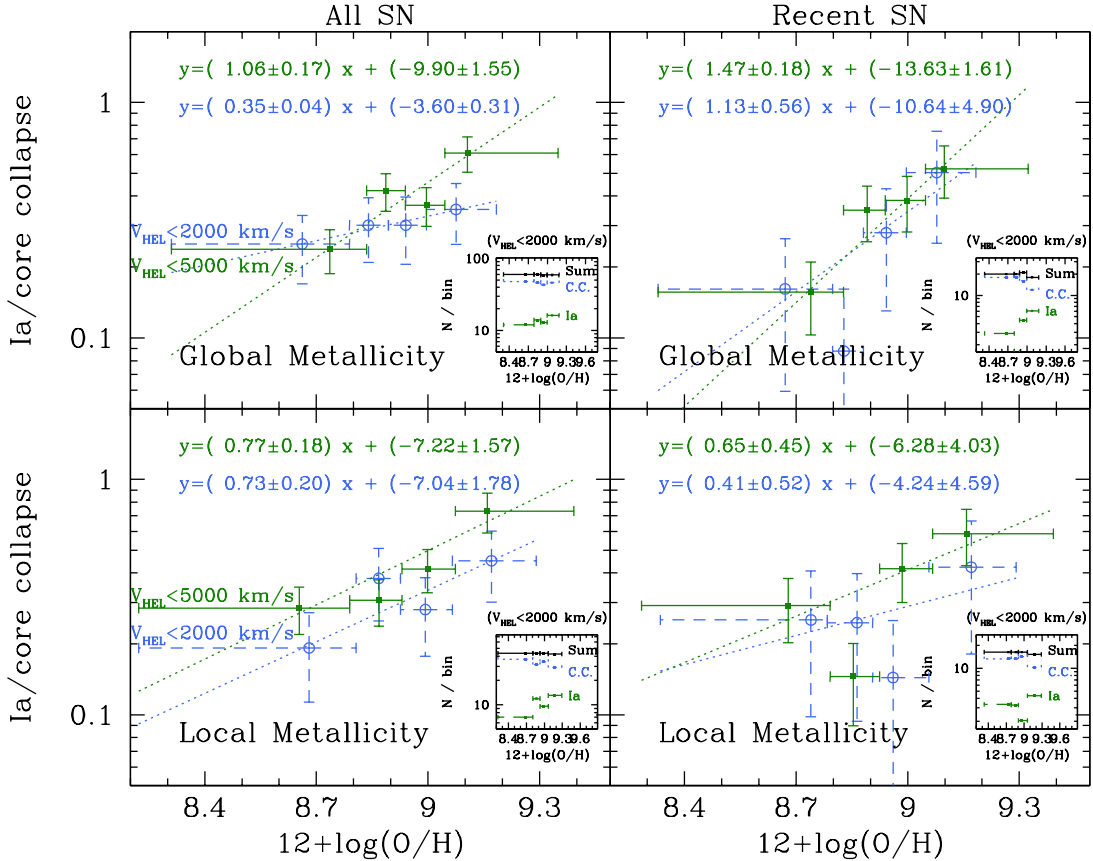}
  \caption{Number ratio of SNIa/CCSN as a function of global
    {\it(top)} and local ({\it bottom}) galaxian metallicity. Data are
    displayed (left panels) for our two samples adopting various
    maximal heliocentric velocities V$_{HEL}$ of the host
    galaxies: 2000 km/s ({\it dashed}), and 5000 km/s ({\it solid}).
    The same is shown in the right panels but keeping only recent SN
    (events that occured during and after 2000).}
  \label{FigMetIa} 
\end{figure*}
% 
%______________________________________________________________

Finally, our results for the SNIa/CCSN ratio are plotted in 
Fig. \ref{FigMetIa} as
a function of global (top) and local (bottom) metallicities.  Results
are plotted for the two samples with values of the maximal
heliocentric velocities of the host galaxies V$_{HEL}<$ 2000, and 5000
km/s. The left and right panels show respectively the results using
all the SN, or only the recent ones.  It is clearly seen that:

i) For global metallicities, the slope of the relation differs in the
two sample defined by different maximal velocity by quite a large
amount. This reflects the difference mentioned in section 3.2 for the
N(Ia)/N(CC) ratio as a function of the magnitude.
We note that if we restrict ourselves to recent SN (right part of Fig.
\ref{FigMetIa}), the data from the 2 samples (V$_{HEL}<$2000, and 5000 km/s) get
within error-bars from each-other (these error-bars are however
large). This suggests misclassifications do play a role in our
N(Ia)/N(CC) ratio and the slope of the relation is quite uncertain.
However, in any cases, we do find a correlation.

ii) The same tendency is obtained for local metallicities.  We note
that, although the number of SNIa in each bin is rather small (see
inserts for the case of V$_{HEL}<$2 000 km/s) the lifetimes of the
progenitors of SNIa are quite long, in general, and thus we have not
the problem described in the previous paragraphs for the SNIc/SNIb
ratio (the more so, since the numbers of CCSN in each bin are quite
substantial). Here again, very similar trends are obtained restricting
ourselves to recent SN (even if large statistical error-bars make the
slope less significant, especially for the V$_{HEL}<$2000 km/s
sample). The variation of the ratio within the whole range of
metallicity is larger than the uncertainties (mentioned in section 2)
due to the \cite{shaw79} effect, or the presence of sub-luminous SN.
We conclude then that the trend of increasing SNIa/CCSN ratio
  with metallicity is likely to be real, although the relation between
  the SNIa/CCSN ratio and metallicity still carries large
  uncertainties.

\section{Discussion}
\subsection{The  $N(Ibc)/N(II)$ ratio }

\begin{figure}
   \centering
 \includegraphics[width=8.5cm]{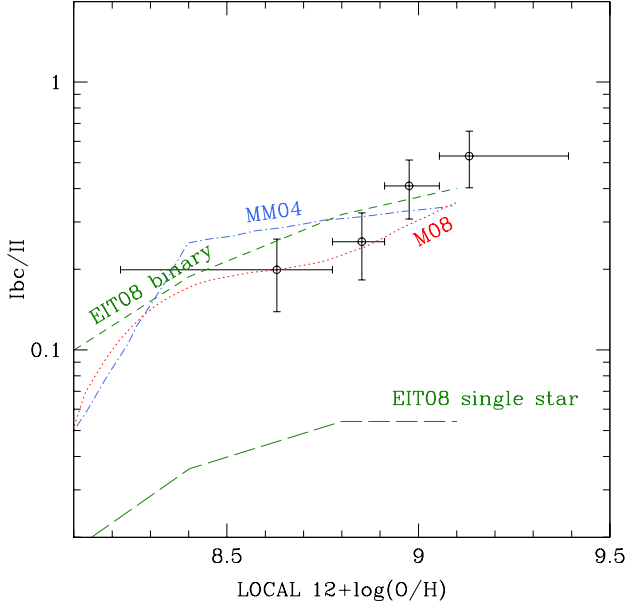}
 \caption{Number ratio of SNIbc/SNII as a function of local galaxian
   metallicity and comparison to single star rotating models of Maeder
   and Meynet (2004, MM04, {\it dot-dashed}) and of Meynet et al.
   (2008, M08, {\it dotted}), single non-rotating stellar models of
   Eldridge et al.  (2008, EIT08, {\it long dashed}) and binary
   non-rotating stellar models of Eldridge et al.  (2008, EIT08, {\it
     short dashed}).  }
         \label{Fig:Model_comparison}
   \end{figure}

The variation of $N(Ibc)/N(II)$ ratio with metallicity is quite well
established now (PB03, Prieto et al. 2008 and this work). Equally well
established is the fact that non-rotating single star models can
reproduce neither the observed trend nor the average value of that
ratio. This is clearly seen in Fig.  \ref{Fig:Model_comparison}, 
where recent resuts of
Eldridge et al.  (2008) are plotted ({\it long dashed curve}).

As suggested in PB03, single star models with rotation are promising in that respect. 
In Fig. \ref{Fig:Model_comparison}  we present results of such
models from the Geneva group (Maeder and Meynet 2004, {\it dot-dashed curve}; Meynet et al. 2008, {\it dotted curve}).
They are both obtained for a power-law IMF with slope x=-1.35. Their behaviour is compatible with
observational data, especially if  observational error bars are taken into account.

An even better fit to the data is obtained by the recent binary star
models of Eldridge et al. (2008). This result is somewhat surprising,
since it is not {\it a priori} obvious how metallicity can affect to
such extent the evolution of stars in binaries. Eldridge et al. (2008)
argue that high metallicity favors a more extended envelope for
massive stars and, in the case of close binary systems, more extended
envelopes make easier the loss of mass through Roche lobe overflow
into the secondary.

In view of those results, it appears difficult to decide whether
stellar rotation or binary evolution is at the origin of the observed
trend. Theoretical uncertainties are quite important in both cases (but
certainly more in the case of binary evolution), making it premature
to draw firm conclusions. It may well be that both factors contribute
to the observed trend.

Stellar models predict the metallicity dependence of M$_{Ibc}$ (the
minimum mass for a single star to lose its hydrogen envelope); then,
folding with a stellar MF allows one to calculate the resulting
$N(Ibc)/N(II)$ ratio vs. metallicity, as e.g.  in Fig. \ref{Fig:Model_comparison}. 
Inversely,
observed $N(Ibc)/N(II)$ ratio vs. metallicity can be used to evaluate
M$_{Ibc}$ vs. metallicity. This was done for the first time in PB03,
who predicted the metallicity dependence of M$_{Ibc}$ on the basis of
then available data for $N(Ibc)/N(II)$ vs. metallicity.  In Fig. \ref{Fig:LimMassMetals} 
we
present the result of PB03 (thin solid curve) and of our new
evaluation (thick solid curve) for a slope x=-1.35 of the IMF. The two
curves are close to each other and not very different from the
theoretical predictions of Meynet et al. (2008, {\it dotted curve} )
at low metallicities. Notice that, in order to compare the results of
Meynet et al. (2008), expressed as a function of Z/\zs, to our own
which are expressed as a function of O/H, we assume that the solar
oxygen abundance is log(O/H)$_{\odot}$+12=8.8; this value is close to
the one determined recently for the Sun
log(O/H)$_{\odot}$+12=8.76$\pm$0.007 (Caffau et al. 2008).  At high
metallicities observationally determined M$_{Ibc}$ is systematically
lower than the predictions of Meynet et al. (2008), \ but by only a
couple of solar masses. Notice that the theoretical predictions of
M$_{Ibc}$ are independent of the IMF, while the empirical
determinations do depend on it: a steeper IMF would produce a
M$_{Ibc}$ vs. metallicity curve lower by a few solar masses than the
one shown in Fig. \ref{Fig:LimMassMetals}. 
Inversely, the 
theoretical $N(Ibc)/N(II)$ ratio
vs. metallicity does depend on the IMF: a steeper IMF would produce
$N(Ibc)/N(II)$ ratios lower than depicted in Fig. \ref{Fig:Model_comparison}.

Given the various uncertainties, our estimates for the minimum mass of
SNIbc at solar metallicity are in fairly good agreement with empirical
estimates for the minimum mass of WN stars in the Milky Way, which
lies in the 20-25 Msun range (Massey et al. 2001, Massey 2003,
Crowther 2007).

%
%_________________________

\subsection{The  $N(Ic)/N(Ib)$ ratio }

In the case of $N(Ic)/N(Ib)$ ratio, the observational situation is not clear at present, since different trends are obtained
as a function of global and local metallicity; we argued in the previous section that small number statistics are at the
origin of this dichotomy. One may only determine a global value of  $N(Ic)/N(Ib)\sim$1.6, i.e. there are about 50\% more
SNIc than SNIb. Fryer et al. (2007) find that this high ratio is `` ... against intuition in the single star case and it may be a further
argument in favor of binary origin for SNIc''.  However, a simple evaluation of the limiting masses on the basis of global
observations and for a Salpeter IMF with slope x=-1.35 gives M$_{Ib}$=24 \ms \  and 
M$_{Ic}$=31 \ms \ (Arbutina 2007), i.e. it is 
easy to obtain the observed ratio if it is assumed that SNIb originate from a  limited range of masses (between
24 and 31 \ms, with more massive stars exploding as SNIc).

\begin{figure}
  \centering
  \includegraphics[width=8.5cm]{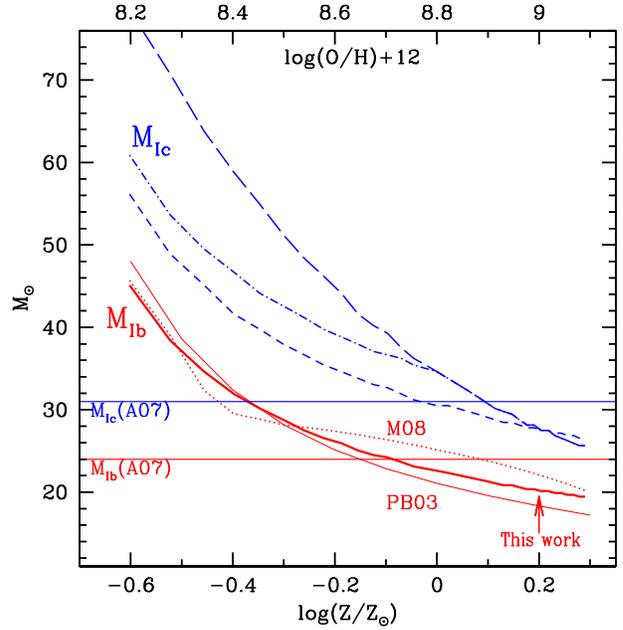}
  \caption{Minimum mass for single stars becoming SNIb or SNIc as a
    function of metallicity, according to theoretical and empirical
    evaluations. The two {\it horizontal lines} are from Arbutina
    (2007), based on measured global SNIbc/CCSN and SNIc/SNIb ratios.
    The three lower curves are for M$_{Ib}$, from theoretical (Meynet
    et al. 2008, {\it dotted}) and empirical evaluations (PB03, {\it
      thin solid} and this work {\it thick solid}). Finally, the three
    upper curves are for M$_{Ic}$ from this work, {\it assuming} that
    the M$_{Ib}$ vs metallicity curve of this work is correct and that
    the IMF extends up to 100 M$_{\odot}$. The three curves are
    obtained then assuming that: i) the SNIc/SNIb ratio is 1.6, as in
    upper left panel in Fig. \ref{FigMetcsurb} ({\it short dashed}), ii) the
    SNIc/SNIb ratio depends on metallicity as in lower left panel in
    Fig.  \ref{FigMetcsurb} extrapolated below the lowest observed metallicity ({\it
      long dashed}), and iii) the SNIc/SNIb ratio depends on
    metallicity only for $Z>$Z$_{\odot}$ and remains equal to unity
    below solar ({\it dot-dashed}) .  The two metallicity scales match
    each other for log(O/H)$_{\odot}$+12=8.8.  }
  \label{Fig:LimMassMetals}
\end{figure}

From the theoretical point of view, there is a dearth of predictions
for $N(Ic)/N(Ib)$ even for the single star case. It is generally
assumed that SNIc originate from more massive stars than SNIb.
However, there is no agreement on the amount of He left in the
envelope for a star to explode as SNIc. Moreover, although it is
generally agreed that the most massive stars end up in black holes, it
is not clear what happens when a black hole is formed: is a bright
optical display still obtained, or is the explosion underluminous or
even failed (see e.g.  Fryer et al. 2007 and references therein)? In
the latter case, the upper part of the IMF would not contribute to
SNIc and the $N(Ic)/N(Ib)$ ratio could be small.

For illustration purposes we present in Fig. \ref{Fig:LimMassMetals} three curves for the
mass limit $M_{Ic}$ as a function of metallicity, on the basis of
various assumptions about the true observational trend.  We assume
that the stellar IMF extends up to 100 \ms{} and that the results of
this work for M$_{Ib}$ ({\it thick solid curve} in Fig. \ref{Fig:LimMassMetals}) are
correct.  The slope of the IMF plays little role in the resulting
$M_{Ic}$ and we shall consider here only the case of x=-1.35 (Salpeter
slope). We proceed by considering three possible cases for the
variation of $N(Ic)/N(Ib)$ ratio with metallicity, on the basis of the
results presented in Fig. \ref{FigMetcsurb}.

%-----------------------------------------------------------
   \begin{figure*}
   \centering
 \includegraphics[angle=-90,width=0.8\textwidth]{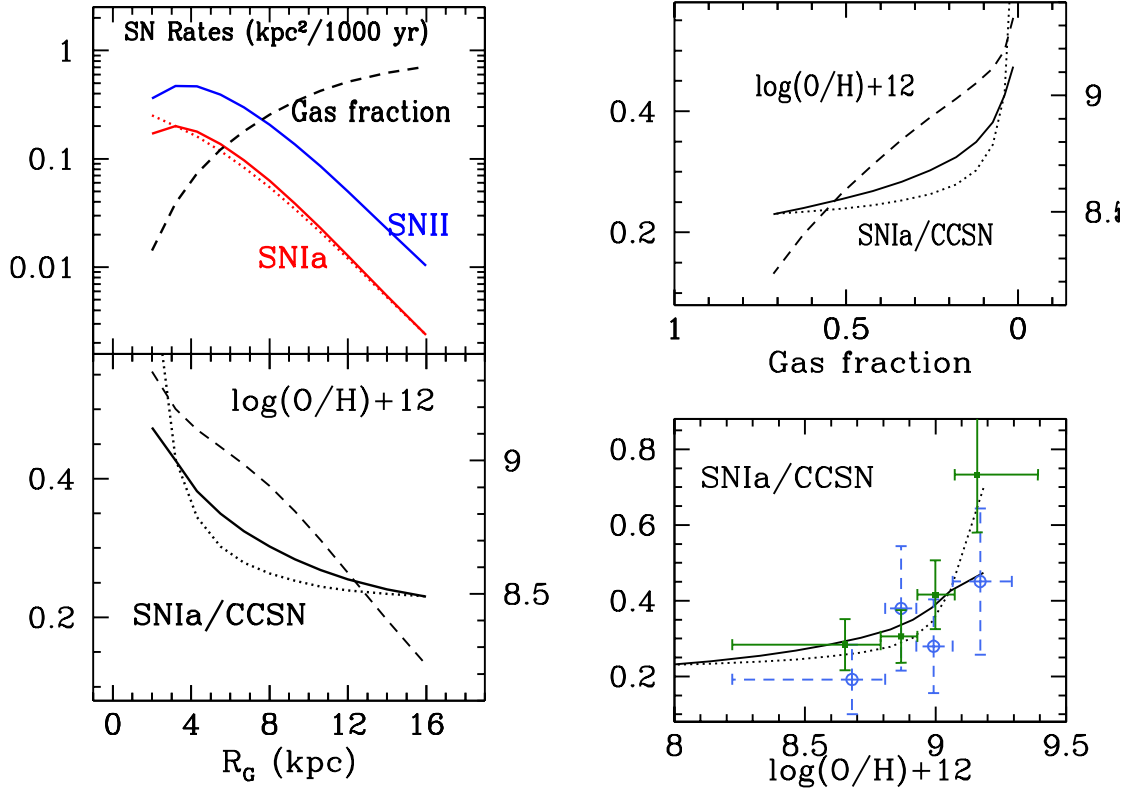}
 \caption{Illustration of the variation of SNIa/CCSN ratio as a
   function of metallicity, with a realistic model of galactic
   evolution (from Boissier and Prantzos 1999). {\it Top left}: Rates
   of CCSN ({\it solid}) and SNIa ({\it solid:} model, {\it dotted:}
   analytical) and gas fraction ({\it dashed}) as a function of
   galactocentric radius. {\it Bottom left}: SNIa/CCSN ratio ({\it
     solid:} model, {\it dotted:} analytical) and metallicity profile
   ({\it dashed}, right vertical axis). {\it Top right}: SNIa/CCSN
   ratio ({\it solid:} model, {\it dotted:} analytical) and
   metallicity ({\it dashed}, right vertical axis). {\it Bottom right}:
   SNIa/CCSN ratio ({\it solid:} model, {\it dotted:} analytical) as a
   function of metallicity; comparison is made to the data for local
   metallicity of Fig. 8 ({\it bottom-left}).  }
         \label{FigMW}
    \end{figure*}

i) $N(Ic)/N(Ib)$=constant = 1.6 (upper left panel in Fig. \ref{FigMetcsurb}). This
leads to the lowest of the three curves for M$_{Ic}$ in Fig. \ref{Fig:LimMassMetals} ({\it
  short-dashed curve}): the curve runs almost parallel to the one for
M$_{Ib}$, at a ``distance'' of a 6-10 \ms, i.e. SNIb are produced only
for a limited range of stellar masses running from 45 to 55 \ms \ at
the lowest metallicities and from 20 to 26 \ms \ at the highest
metallicities.

ii) $N(Ic)/N(Ib)$=Z/\zs \ at all metallicities (lower left panel of
Fig. \ref{FigMetcsurb}, trend extrapolated to lower than solar metallicities). In that
case we obtain the highest lying curve in Fig. \ref{Fig:LimMassMetals} ({\it long-dashed}).
At low metallicities the ratio is small and the limiting mass M$_{Ic}$
is as high as 75 \ms, while at high metallicities we recover the
results of the previous case.

iii) $N(Ic)/N(Ib)$=Z/\zs \ at Z$>$\zs \ and $N(Ic)/N(Ib)$=1 for
Z$<$\zs . This case leads to a curve intermediate between the two
previous ones ({\it dot-dashed} in Fig. \ref{Fig:LimMassMetals})

The obtained curves confirm the finding of Arbutina (2007) on the
basis of global SN ratios, namely that it is possible to have both
SNIb and SNIc solely from single star evolution; this suggests that
the concern expressed in Fryer et al. (207) was unfounded.  If the
$N(Ic)/N(Ib)$ ratio is $\sim$2 at high metallicities, then the range
of masses producing SNIb is rather limited. This is also found in the
single star models with rotation of Meynet et al. (2008), although no
quantitative predictions are given.  The wide spacing between the
curves of M$_{Ic}$ corersponding to cases (i) and (iii) in Fig. \ref{Fig:LimMassMetals}
suggests that a substantial effort is required in order to pin down
the true metallicity dependence of $N(Ic)/N(Ib)$ ratio through better
statistics.

\subsection{The  $N(Ia)/N(CC)$ ratio }

   In Sec. 3.2 we gave an explanation of the observed variation of
   $N(Ia)/N(CC)$ ratio with global metallicity. Here we provide a
   similar argument for the observed variation of $N(Ia)/N(CC)$ ratio
   with local metallicity. In the case of a star forming galactic
   disk, Eq. (5) can be rewritten in terms of local surface densities
   of stars $\Sigma_*(R)$ and of star formation rate $\Psi(R) \propto
   \Sigma_g^k(R)$, where $\Sigma_g(R)$ is the gas surface density at
   galactocentric distance $R$ and $k$=1.4 the coefficient in the
   Kennicut (1998) empirical ``star formation law''.
\begin{equation}
{{N(Ia)}\over{N(CC)}} \ = \ A  + \ B \ {{\Sigma_*}\over{\Sigma_g^k}}
\end{equation}
where all variables depend on radial distance $R$. Stellar profiles in
galactic disks are usually fitted with exponentials of scalelength
$R_*$, i.e. $\Sigma_*(R)\propto {\rm exp}(-R/R_*)$. Corresponding
gaseous profiles are always much flatter than stellar ones and if fitted by exponentials they
would be $\Sigma_g(R)\propto {\rm exp}(-R/R_g)$ with $R_g > 2 R_*$. In the case of the Milky Way, for instance,
one has $R_*\sim$2.5 kpc and $R_g\sim$8 kpc; further  examples for external disk galaxies can be found in
e.g. Boissier et al. (2003). Eq. (7)  is then rewritten as
\begin{equation}
{{N(Ia)}\over{N(CC)}} \ = \ A  + \ C \ {\rm exp}(-{{R_g - k R_*}\over{R_g R_*}} \ R)
\end{equation}
where $C$ is a new constant. It is clearly seen that, for reasonable values of $k$ ($<$2), the expression
(8) is a decreasing function of radius $R$.  Thus, in galactic disks
it is expected that the $N(Ia)/N(CC)$ ratio will increase towards the
inner galaxy, i.e. it will be correlated to metallicity, as observed.

We illustrate this behaviour in Fig. \ref{FigMW}, where we plot the relevant
quantities for the case of the Milky Way disk.  All curves are
obtained from an updated successfull model of the Galactic disk (from
Boissier and Prantzos 1999). For the SNIa rate, the formalism by
Greggio and Renzini (1983) is adopted in the model and the
corresponding results are displayed with {\it solid} curves in all
panels. We also apply the simple analytic expression of Eq. (8) and
the corresponding results are plotted with {\it dotted} curves. Fig.
\ref{FigMW} displays the present-day (T=12.5 Gyr) radial profiles of CCSN, SNIa
and gas fraction (top left panels), the profiles of $N(Ia)/N(CC)$
ratio and oxygen (bottom left panels) and shows clearly that both
$N(Ia)/N(CC)$ and metallicity increase at lower gas fractions, i.e. in
the inner disk (top right panel). Finally, in the bottom right panel
we show that the resulting $N(Ia)/N(CC)$ vs metallicity relation
compares favorably with the data of Fig. \ref{FigMetIa} ; the analytical
prescription for the SNIa rate leads to a more steeply rising
$N(Ia)/N(CC)$ ratio with metallicity than the prescription of Greggio
and Renzini (1983), but in both cases the agreement with observations
is satisfactory. Obviously, an increase of the observational sample of
SN will allow in the future to reduce error bars and to constrain
prescriptions for the SNIa rate.

\section{Summary}

In this work we derive relationships between ratios of various SN
types and metallicity of host galaxies. For that purpose we detemine
either global metallicities (reflecting the composition at radius
$R=$0.4 $R_{25}$) or local ones, i.e.  at the position of the SN
inside the host galaxy. In the former case we use the well known
metallicity-magnitude relationship, a technique applied already in
PB03 (albeit with a smaller SN sample). In the latter case, we use the
fact that galaxian metallicity gradients appear to be $\sim$constant
when expressed in dex/$R_{25}$; this method is applied for the first
time, to our knowledge, to the determination of local metallicities in
disks and appears quite promising.  We made a number of tests,
defining several samples and using different values for the abundance
gradients, in order to make sure the observed trends are not biased.

We find that $N(Ibc)/N(II)$ ratio increases with both global
metallicity (as already found in PB03 and in Prieto et al. 2008) and
with local one. Our study reduces considerably error bars of previous
works. We consider this result as established now (the variation of
$N(Ibc)/N(II)$ with metallicity is larger than the changes that could
cause any biases we can think of, and the trend is consistently
obtained for almost all our samples, e.g. with various assumptions or
limits on $V_{HEL}$, with the exception of the $V_{HEL}<$ 2000
  km/s sample in which the ratio for the higher {\it global} metallicity bin is
  lower than expected for such a trend) and we discuss it in terms
of either single star models with rotation or binary evolution models.
In view of observational and theoretical uncertainties (certainly
larger in the case of binary evolution than in the case of single
stars) we find it difficult to chose between the two possibilities.
Assuming that only single stars produce SNIbc, we derive the empirical
$M_{Ibc}$ vs.  metallicity relation and we find it to be compatible
with the one obtained with the latest models of the Geneva group
(Meynet et al.  2008).

We study the $N(Ic)/N(Ib)$ ratio and we find it consistent with being
constant w.r.t. global metallicity but increasing with local one. We
attribute this difference to small number statistics and we believe
that larger SN samples in the future will allow to distinguish between
the two possibilities. Nevertheless, we derive the empirical
$M_{Ic}$=f(Z) relation assuming again that only single stars are at
the origin of SNIb and SNIc. We find that this assumption leads to
SNIb being produced within a relatively limited range of stellar
masses, as found in rotating star models of Meynet et al. (2008). We
show that the single star channel can justify $N(Ic)/N(Ib)$ ratios as
high as 2 and we disagree in that respect with the concern expressed
in Fryer et al.  (2007), namely that such a high ratio favours the
binary channel for SNIc.

Finally, we find for the fist time an unexpected corelation between
the ratio of $N(Ia)/N(CC)$ and metallicity, both global and local.
Although the precise values of the slope and intersect are subject to
large uncertainties (various samples provide different results), the
trend is statistically secure (the null hypothesis has a probability
of only 7\% using local metallicity and the conservative $V_{HEL}<$
2000 km/s sample).
%%%%%%%%%
We argue that this is not a causal relationship, contrary to the
previous cases; instead, both $N(Ia)/N(CC)$ and metallicity are higher
in regions of smaller gas fractions (or lower specific star formation
rates). We develop this argument analytically and we illustrate it
with a quantitative application to the case of the Milky Way disk.

Comparing to the situation only $\sim$ 6 years ago, we find that the
increase of the SN sample size since PB03, and the use of local
metallicities rather than global ones allowed us to establish
%beyond any reasonable doubt
the strong likelihood of 
the $N(Ibc)/N(II)$ vs
metallicity trend. We expect then that a similar increase in the
future will allow one to establish the $N(Ic)/N(Ib)$ vs. metallicity
trend and to probe with greater accuracy the intricacies of the SNIa
rate.

\begin{acknowledgements}

  We thank S. Basa and T. Zhang for useful discussions. We thank the
  referee, S. Smartt for his detailed review of the paper and his
  suggestions. We acknowledge the usage of the HyperLeda database
  (http://leda.univ-lyon1.fr), as well of the Padova-Asiago supernova
  catalogue, from the Padova-Asiago Supernova Group
  (http://web.pd.astro.it/supern/)

\end{acknowledgements}

\end{document}